\documentclass[reprint,aps,amsmath,amssymb,nofootinbib]{revtex4-1}
\usepackage{amssymb}
\usepackage{amssymb}
\usepackage{amssymb}
\usepackage{graphicx}
\usepackage{graphics}
\usepackage{amsmath}
\usepackage{placeins}
\usepackage{hyperref}
\usepackage[dvipsnames]{xcolor}
\usepackage{wasysym}
\usepackage{soul}
\usepackage{float}
\usepackage{array}
\usepackage{tabulary}
\usepackage{caption}
\usepackage{fancyhdr}
\usepackage{epstopdf}
\usepackage{multirow}

\begin{document}
	
	\title{High Gradient Testing of Off-Axis Coupled C-band Cu and CuAg Accelerating Structures }
	
	\author{Mitchell Schneider}
	\email{mitchs@slac.stanford.edu}
	\author{Valery Dolgashev}
	\author{John W. Lewellen}
	\author{Sami G. Tantawi}
	\author{Emilio A. Nanni}
	\affiliation{SLAC National Accelerator Laboratory, Stanford University, Menlo Park, CA, U.S.A}
	
    \author{Muhammed Zuboraj}
    \email{zuboraj@lanl.gov}
    \author{Ryan Fleming}
    \author{Dmitry Gorelov}
    \author{Mark Middendorf}
    \author{Evgenya I. Simakov}
    
    \affiliation{Los Alamos National Laboratory, Los Alamos, NM, U.S.A.}
    \collaboration{Both M. Schneider and M. Zuboraj Contributed Equally}
	
\begin{abstract}
We report the high gradient testing results of two single cell off-axis coupled standing wave accelerating structures. Two brazed standing wave off-axis coupled structures with the same geometry were tested: one made of pure copper (Cu), and one made of a copper-silver (CuAg) alloy with a silver concentration of 0.08\%. A peak surface electric field of 450 MV/m was achieved in the CuAg structure for a klystron input power of 14.5 MW and a 1 $\mu$s pulse length, which was 25\% higher than the peak surface electric field achieved in the Cu structure. The superb high gradient performance was achieved because of the two major optimizations in the cavity’s geometry: 1) the shunt impedance of the cavity was maximized for a peak surface electric field to accelerating gradient ratio of $\sim$2 for a fully relativistic particle, 2) the peak magnetic field enhancement due to the input coupler was minimized to limit pulse heating.  These tests allow us to conclude that C-band accelerating structures can operate at peak fields similar to those at higher frequencies while providing a larger beam iris for improved beam transport.
\end{abstract}

\maketitle

Reducing the overall physical footprint of a particle accelerator has become important for many accelerator applications in industry, medicine, national security, and basic sciences \cite{FEL,lclsHE,Nextgenlinac,C3}. In response, there has been a concerted effort to develop high gradient accelerating structures which allow achieving the required beam energy within a shorter length. These compact accelerators can be used for many applications such as high brightness light sources \cite{FEL,lclsHE}, high energy facilities \cite{Nextgenlinac,bai2021c,C3}, medical radiotherapy \cite{SbandXlu,Medlinac} and industrial LINACs \cite{Industry}. The high gradient accelerators must often meet a user’s requirement of transporting a high charge or a high current particle beam which restricts the size of the minimum aperture of the accelerating structures and makes it preferable to operate at lower microwave frequencies. In many previous works, high gradient operation of X-band (11.424-11.992 GHz) accelerating structures was successfully demonstrated and studied \cite{HGjenya}. Here we explore development of high gradient accelerators at C-band (5.712 GHz) where the beam aperture can be twice as large as at X-band, reducing the level of higher order modes that can be excited by the accelerating particle beam.

The C-band cavities tested in this project were optimized as standing wave accelerating structures to be used with distributed coupling topologies \cite{SamiDC}. However, this cavity’s geometry can be utilized for accelerating multiple species of particles by adjusting the phase advance between subsequent cells \cite{SbandXlu,C3,SamiDC,SRFPF}. For example, fully relativistic electrons or protons would have a $\sim$100$^o$ phase advance/cell in this structure, and $\beta$=~0.5 protons would accelerate at a 180$^o$ phase advance/cell. 

The important figures of merit for performance of any high gradient structure are the achievable peak surface fields, accelerating gradient and subsequent breakdown rate (BDR) at the given field \cite{HGjenya,GeoValery,PHValery,SRFValery,othman2020experimental}. A breakdown is a vacuum arc discharge inside of the structure which generates an excursion of gases, particulates, and ions from the surface. Radiofrequency (RF) breakdowns are related to multiple phenomena including pulse heating and field emission/dark current \cite{PHValery,Jiahang}. The BDR is defined as the probability of a breakdown event per a RF pulse normalized to the length of the accelerating structure for a given RF pulse length.

C-band accelerating structures have been previously studied by multiple institutions and projects, such as SwissFEL \cite{SwissFEL}, SINAP \cite{SINAP}, SPARC LAB \cite{SPARC}, the FEL Spring-8 \cite{Spring8}, and the Korean National Fusion Research Institute (KNFRI) \cite{KNFRI}. All previous projects used traveling-wave C-band structures with exception of KNFRI \cite{KNFRI} that used standing-wave structures. The peak surface fields in those C-band structures were in the range of 80-150 MV/m while requiring input power from the klystron on the order of 10s MW. The breakdown rates in these multi-cell accelerating structures varied between 1$\times10^{-5}$ and 1$\times10^{-6}$ (1/pulse/meter) for the pulse length in the range of 0.5-1 µs. 

\begin{figure}
\begin{center}
\includegraphics[width=8.6cm]{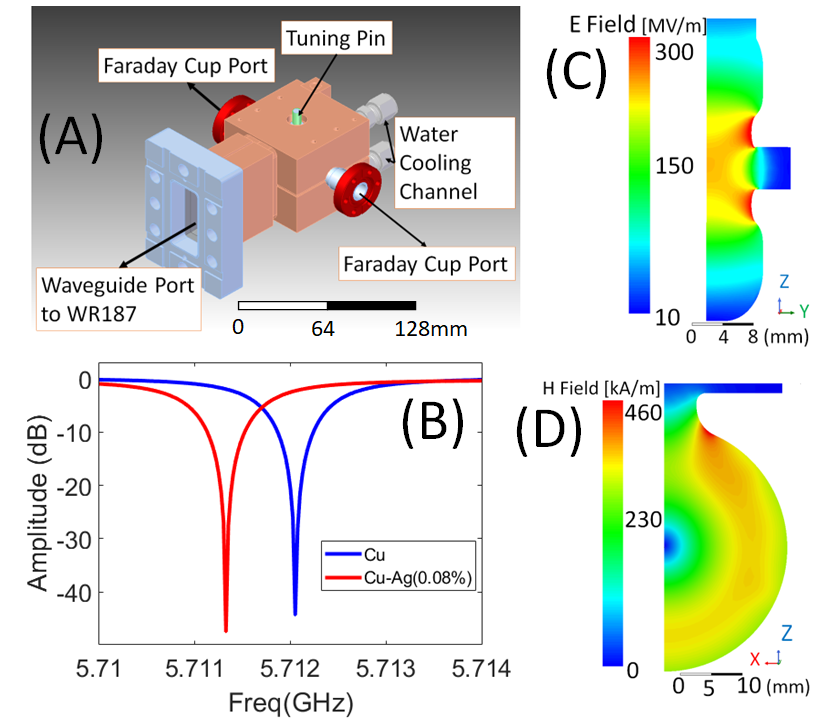}
\caption{ (a) The mechanical CAD Modal of the 5.712 GHz copper cavity, (b) the reflection coefficients measured during cold-testing of two fabricated cavities, one made of copper and the other one of copper-silver. The results of HFSS computations for (c) the electric field plotted at the central cross-section the cavity and (d) the magnetic field on the surface of the cavity. The dashed lines show the symmetry axis field are for 4MW of power dissipated in the cavity.}
\label{HFSSmod}
\end{center}
\end{figure}

High gradient testing of accelerating structures operating at X-band is routinely conducted at SLAC National Accelerator Laboratory demonstrating peak surface electric fields greater than 350 MV/m for 1$\times10^{-4}$ to 1$\times10^{-2}$ BDR(1/pulse/meter) for RF pulse length of 85-300 ns \cite{GeoValery,PHValery,SRFValery}. This is similar with testing of X-band structures at CERN/KEK \cite{CERNKEK} and S-band work form CERN \cite{CERN} with both showing a maximum achievable surface electric field of ~225 MV/m for a  BDR of less than 1$\times10^{-7}$(1/pulse/meter) for the 250 ns pulse length at X-band and surface field of ~220 MV/m for a  BDR of less than 1$\times10^{-6}$ (1/pulse/meter) for the 1.2 $\mu$s pulse length at S-band. Those same X-band (11.424 GHz) experiments showed that the breakdown rate correlates with peak pulse surface heating. It is hypothesized that pulsed heating results in cyclic fatigue, defect mobility and physical changes to the surface that enhance surface fields and consequently lead to breakdowns \cite{LANLThm}.
The accelerator cavity made for this project was designed with the goal of maximizing shunt impedance, minimizing the peak surface magnetic field at the input waveguide coupler, and reducing the pulse heating. The structure itself is a C-band version of an S-band (2.856 GHz) accelerating cavity that was designed by X. Lu et al. \cite{SbandXlu} to serve as an energy modulator for $\beta$=0.5 protons utilized in proton radiation therapy. The geometry was scaled down in size to increase the resonant frequency to 5.712 GHz. The two geometric differences are an increased length of the cell to ease machining requirements and the RF feed shape which was tapered due to the relatively smaller dimensions of the WR-187 waveguide compared to the WR-284 waveguide. Fig.~\ref{HFSSmod}a shows a mechanical modal for the cavity. 
The two structures were fabricated for the experiment described. One was made of pure Cu and the other one was made of a CuAg alloy with 0.08\% concentration of Ag to further investigate breakdown rates in CuAg structures compared to Cu structures \cite{SRFValery}. The hypothesis, as proposed in new computational work \cite{LANLThm}, was that adding a small concentration of silver is superior in copper alloy to increase resistance to the thermal and mechanical stresses during an RF pulse which could result in higher achievable fields for the CuAg structure before the onset of breakdown. Fig.~\ref{HFSSmod}b shows the reflection coefficient S11 that were measured during the RF cold tests of the two fabricated cavities.

Fig.~\ref{HFSSmod}c  shows the results of Ansys electronics simulation toolkit HFSS simulations for the magnitude of the electric field plotted at the central plane of the structure. Fig.~\ref{HFSSmod}d shows the results of HFSS simulations for the magnitude of the magnetic field on the surfaces of the cavities and illustrates that the maximum magnetic field is located on the upper walls of the accelerating structure near the input coupler, far away from the beam iris. This is the location where the maximum of peak pulse surface heating will occur during high gradient testing. 
The accelerating parameters of the cavity computed with HFSS can be seen in Table~\ref{T1mod}. Compared to the S-band cavity (2.856 GHz), the scaled C-band cavity has an increased shunt, impedance Rs which is proportional to the square root of the cavity’s frequency, $\sqrt{f}$. Table~\ref{T2mus} shows the Q factors and resonant frequencies extracted from the RF cold test results of the two fabricated cavities with the copper-silver cavity having a slightly higher Q-factor.

\begin{table}
\centering
\caption{Parameters of the C-band cavity as computed by HFSS for a copper structure. For a v=0.5c protons and v=c electrons acceleration.}
\begin{tabular}{ l | c  c  c}

\multirow{2}{*}{Parameter} & Cu & Cu\\
& v=0.5c Proton & v=c Electron  \\ 
\hline
\hline
Length              & \multicolumn{2}{c}{1.58 cm}\\
$a/\lambda$        & \multicolumn{2}{c}{0.0525}\\
Frequency       & \multicolumn{2}{c}{5.712GHz}\\
$\sigma$              &  \multicolumn{2}{c}{58 MS/m}\\
$Q_0$               &  \multicolumn{2}{c}{9762}\\
$Q_{ext}$           &  \multicolumn{2}{c}{10165}\\
\multirow{2}{*}{\centering $R_s$}  & 61.51  & 115.8 \\
&$M\Omega/m$&$M\Omega/m$\\
\multirow{2}{*}{\centering $E_a$}       & 62 MeV/m & 81 MeV/m \\
&$\times\sqrt{P[MW]}$ &$\times\sqrt{P[MW]}$ \\
$E_p/E_a$            & 2.42 & 1.84 \\
$H_p*Z_0/E_a$              & 1.40 & 1.07
\end{tabular}

\label{T1mod}
\end{table} 

\begin{table}
\centering
\caption{Cavity parameters measured in the RF cold test of the Cu and the CuAg cavities.}
\begin{tabular}{ l | c  c  c}
Cold Test Results~ & { Cu} & {CuAg}\\[4pt]
\hline
\hline
Frequency             &  \space\space5.71205 GHz \space\space\space &  5.71133 GHz \\[4pt]

$Q_0$               & 9621 & 9720\\[4pt]
$Q_{ext}$           & 9742 & 9805\\[4pt]
$2*\tau$ &269 ns & 272 ns\\[4pt]
\end{tabular}
\label{T2mus}
\end{table} 

\begin{figure}
\begin{center}
\includegraphics[height=9.cm]{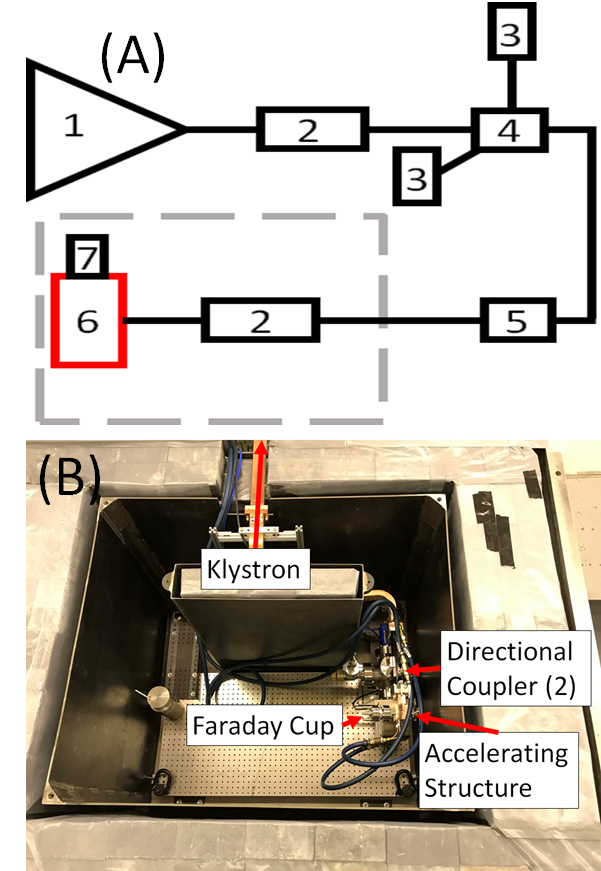}
\caption{(A) The schematic of the RF network at CERF-NM test facility: (1) C-band Canon klystron; (2) high power directional couplers; (3) high power water loads; (4)waveguide 3dB hybrid (Magic T); (5) C-band RF window (D) Cavity under test; (7) Faraday cup; (Dashed line denotes lead box shielding). (B) The photograph of the lead box at CERF-NM with an RF cavity installed for high gradient testing. The breadboard at the bottom of the lead box shielding has one inch spacing between the screw holes. }
\label{CERF}
\end{center}
\end{figure}

The high gradient testing of the two structures was performed at the C-band Engineering Research Facility in New Mexico (CERF-NM) at Los Alamos National Laboratory (LANL)\cite{CERF}. The schematic and a photograph of CERF-NM are shown in Fig.~\ref{CERF}. The facility is built around a 50 MW C-band Canon klystron that can provide up to 25 MW of RF power into the cavity at a 100 Hz repetition rate.

\begin{figure}
\begin{center}
\includegraphics[width=7.cm]{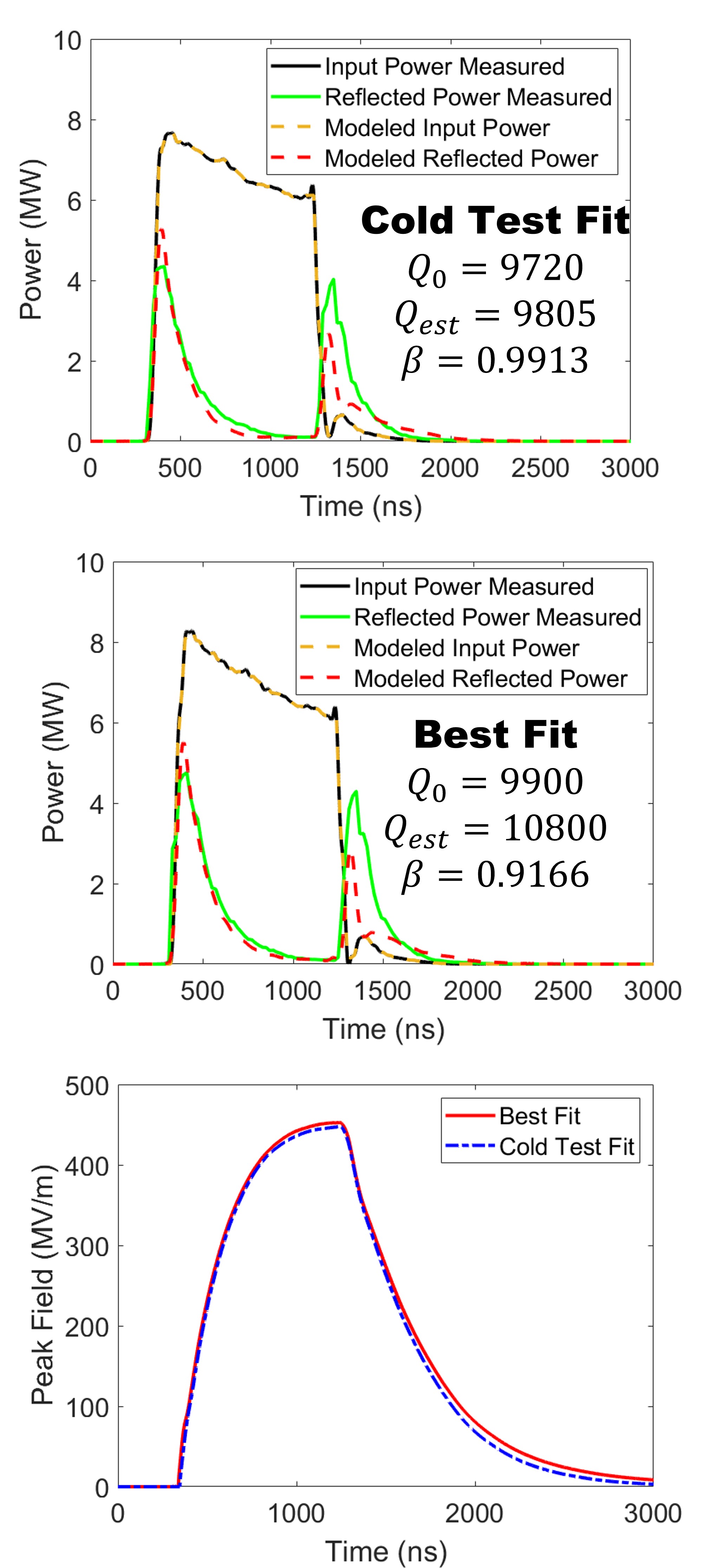}
\caption{Pulse shapes for the forward and reflected power at the cavity entrance measured in the experiment and calculated for the S$_{11}$ parameters that were (a) the cold test fit Q factors and (b) the “best fit” Q factors. (c) The corresponding peak surface field as a function of time for both scenarios. For 8 MW to feed into to the cavity, 14.5 MW was required from the klystron source due to the magic tee in the waveguide configuration (see Fig.~\ref{CERF})}
\label{pulseS}
\end{center}
\end{figure}

\begin{figure*}
\begin{center}
\includegraphics[width=12.cm]{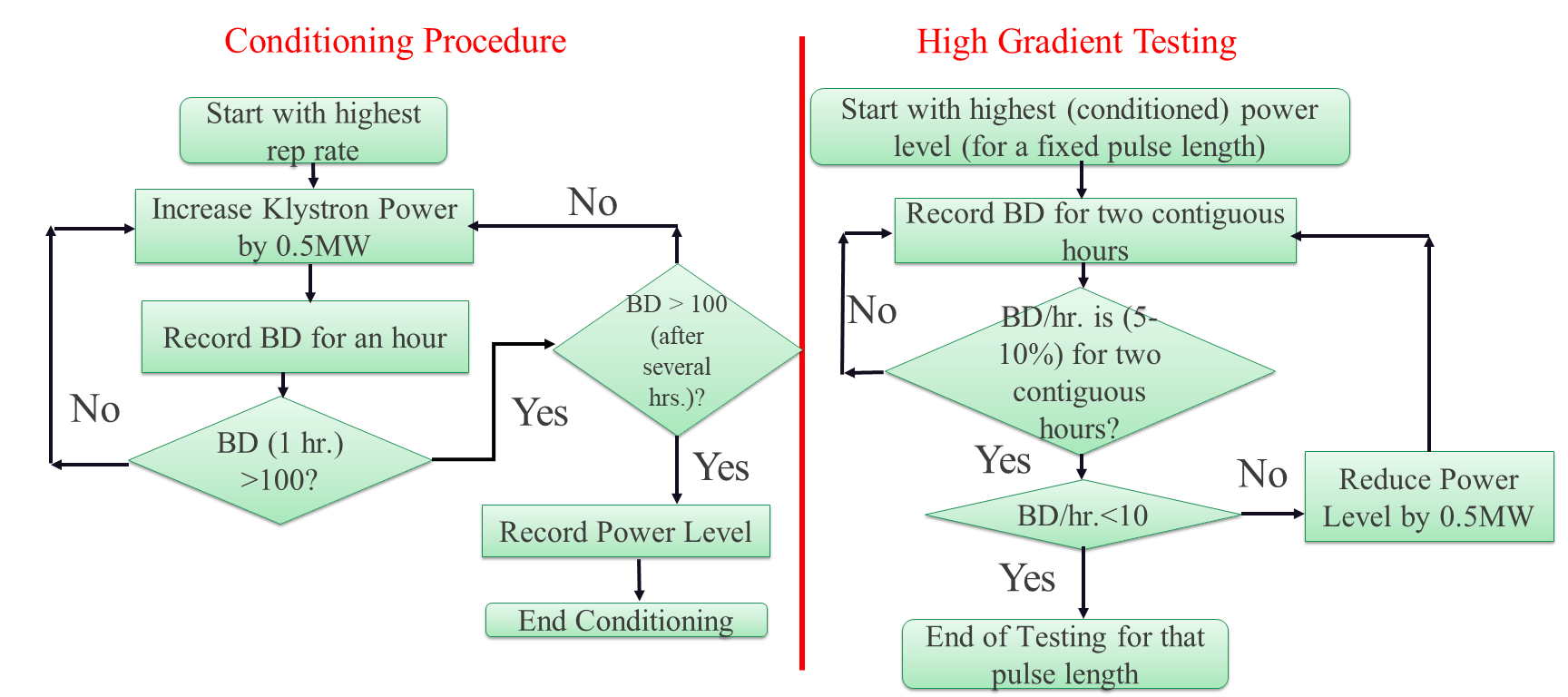}
\caption{Flowcharts of high gradient conditioning procedure at CERF-NM and for mapping of the breakdown rate.Fig.~\ref{CERF}}
\label{FLowchart}
\end{center}
\end{figure*}
\begin{figure}[h]
\begin{center}
\includegraphics[height=12.5cm]{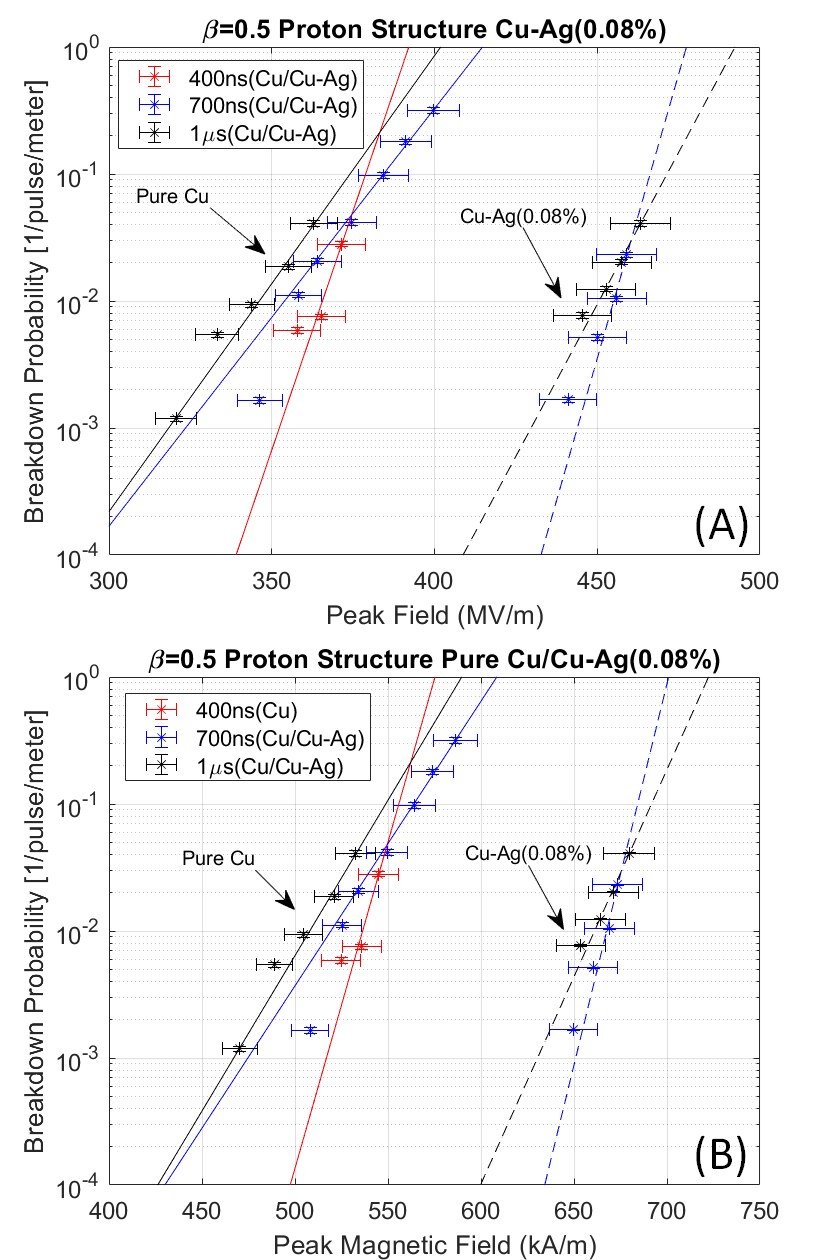}
\caption{BDR plotted (a) as a function of peak electric field ($E_{p}$) and (b) peak magnetic field ($H_{p}$) on the surface of the accelerator structures. The trendlines are the data fitted to an exponential fit. Note that the accuracy of the trendline is low over a large dynamic range.}
\label{BDR}
\end{center}
\end{figure}

During the conditioning process, the forward and reflected pulse shapes were recorded continuously at both directional couplers shown in the schematics in Fig ~\ref{CERF}a. The dark current was monitored using a Kimball Physics FC73a Faraday cup which was mounted on the flange at the beampipe of the cavity. We developed a software package named FEbreak \cite{FEbreak} which continuously monitored and recorded the forward and reflected traces from an analog signal of Ladybug peak power meters (LB480A). The breakdown events were identified by monitoring the Faraday cup signal with the trigger for a breakdown event set to 2.5 mV (twice the noise floor). In the end, the FEbreak software had a pulse capture efficiency of $\geq$95\%, limited by the speed of the data acquisition hardware \cite{FEbreak}. This means that about than 5\% of RF pulses were not captured or analyzed.
To compute the gradient and peak surface fields in the cavity for the given coupled power and pulse shape, the conventional analysis for high gradient normal conducting structures was used based upon a linear equivalent circuit model assuming a constant Q-factor \cite{CalEa}. For this data processing we used the Ohmic and external Q-factors, $Q_{0}$ and $Q_{ext}$, obtained during RF cold testing. During data processing, we found that, there was a difference between the model measured reflected power reflected power. Where the model reflected power decayed faster than  the measured data (see Fig.~\ref{pulseS}a) To correct the model, we used the values of $Q_{0}$=9900 and $Q_{ext}$=10800 for Cu and $Q_{0}$=11120 and $Q_{ext}$=11180 for CuAg which are slightly different from the RF cold testing values in Table.~\ref{T2mus}. The resulting forward and reflected pulses obtained in this “best fit model” are shown in Fig.~\ref{pulseS}b. Fig.~\ref{pulseS}c shows the peak surface electric field in the CuAg cavity calculated as a function of time for the pulse for klystron power of 14.5 MW. Fig.~\ref{pulseS}c shows that making slight adjustment to the cavity’s Q-factors used in the model had a negligible effect on the calculated peak surface fields in the cavity.
Figure~\ref{BDR} shows the collected breakdown statistics for both Cu and Cu-Ag cavities, where the probabilities of RF breakdowns are plotted as functions of peak surface electric field ($E_{p}$), and peak surface magnetic field ($H_{p}$) at different lengths of the RF pulses. The pulse shape for both the Cu and CuAg cavities were of similar shape. It is worth noting that the conditioning process was the same for both the Cu and CuAg cavities, as described in Fig.~\ref{FLowchart}.

It was found that the structure was able to achieve remarkably high gradients with peak surface fields up to 450 MV/m for the CuAg structure and up to 360 MV/m for the pure Cu cavity with breakdown rates in the range of $10^{-1}$ /pulse/meter. The 25\% improvement in the maximum achievable gradient and peak surface magnetic field in the CuAg cavity as compared to the pure Cu cavity as seen in Fig.~\ref{BDR}a agrees with what has been reported previously \cite{SRFValery}.
The results in Fig.~\ref{BDR}a indicate that the BDR data of each structure is clustered by cavity material (overlapping solid lines and overlapping dashed lines). Data was collected for both structures in approximately the same breakdown rate range ($10^{-1}$-$10^{-3}$ /pulse/m). For each cavity, the same BDR at the same peak surface fields are achieved, irrespective of pulse length. This shows that the breakdown statistics are weakly dependent on pulse heating indicating that the mechanism for creating breakdowns in this structure is not dominated by RF pulse heating.

The geometry of this cavity that leads to the high shunt impedance also leads to a low magnetic field on the surface and low peak magnetic to peak electric field ratio. This is similar to an X-band structure that was reported in \cite{GeoValery}. The calculated peak magnetic field BDR plot, shown in Fig. ~\ref{BDR}b, is also pulse length independent, as seen and explained for the peak electric field plots in Fig.~\ref{BDR}  (solid lines for Cu and dashed lines for CuAg). 

\begin{figure}[h]
\begin{center}
\includegraphics[height=12.cm]{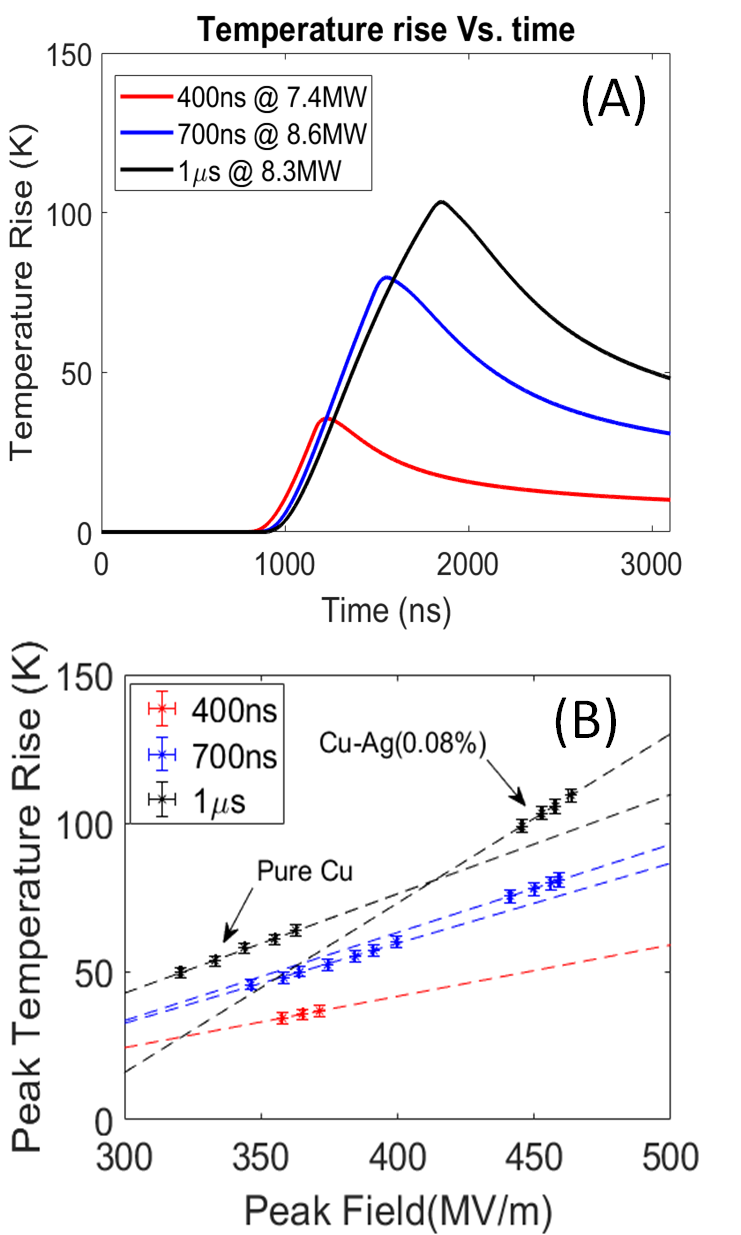}
\caption{BDR plotted (a) as a function of peak electric field ($E_{p}$) and (b) peak magnetic field ($H_{p}$) on the surface of the accelerator structures. The trendlines are the data fitted to an exponential fit. Note that the accuracy of the trendline is low over a large dynamic range.}
\label{PH}
\end{center}
\end{figure}

Fig.~\ref{PH}a shows the evolution of the peak temperature rise inside of the cavity that is due to pulse heating versus time \cite{PHValery}over the duration of the RF pulse. The temperature rises exponentially until it peaks immediately after the RF drive pulse is turned off. It then decays slowly to zero over the 10 ms delay between pulses that come periodically with the repetition rate of 100 Hz. In the experiment, there was no residual temperature rise between pulses see Fig.~\ref{PH}a. Fig.~\ref{PH}b shows that the CuAg structure exhibits higher temperature rise due to pulse heating than its copper counterpart because it conditioned to a higher gradient and the higher peak fields. This structure was still able to achieve higher fields than its copper counterpart because the temperature rises due to pulse heating did not have a significant impact on the breakdown statistics.

\begin{figure}[h]
\begin{center}
\includegraphics[height=7.5cm]{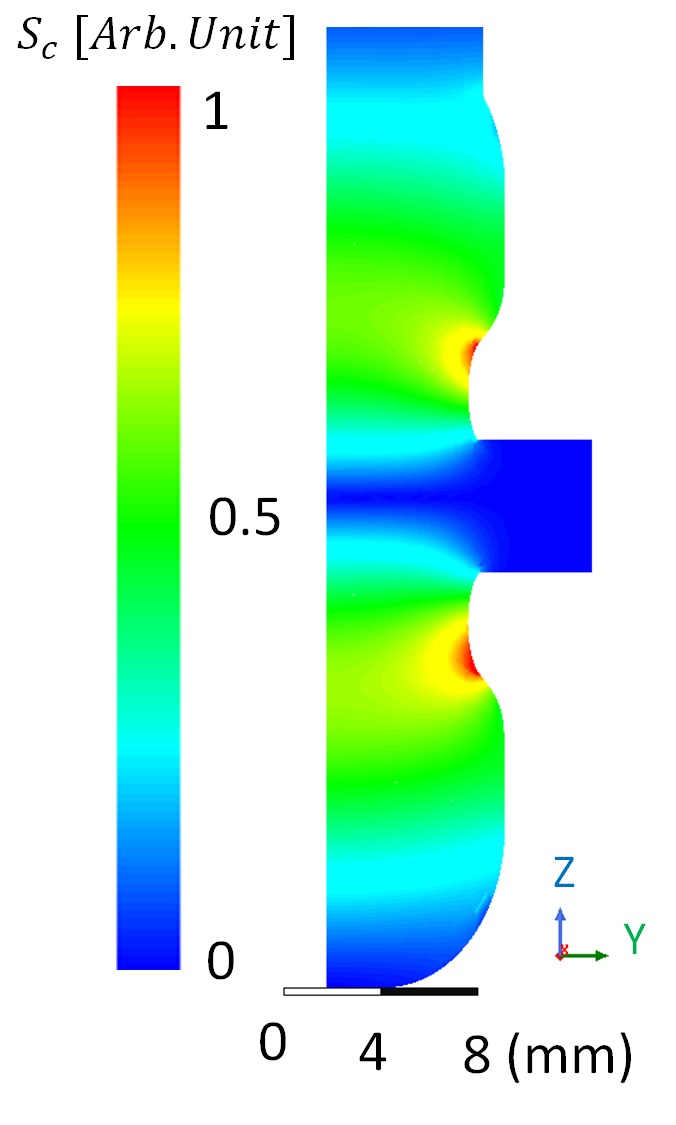}
\caption{$\lvert S_{c}\rvert$|in cross-section.}
\label{Sc}
\end{center}
\end{figure}

Finally, we would like to discuss why this cavity geometry could sustain much higher peak fields than previous C-band accelerating structures. In this investigation, the methodology outlined by A. Grudiev $et~al.$ \cite{grudiev2009new} was used. In paper \cite{grudiev2009new}, extensive testing of X-band accelerating structures showed that the BDR may not be dominated by pulse heating but instead the power coupling between the power flow into a field emission electron source and the RF power inside of the cavity. The authors derived a modified Poynting vector ($S_{c}$) defined:

\[S_c=Re[\vec{S}]+ g_c*Im[\vec{S}]\]

where $S_{c}$ is the standard Poynting vector and $g_{c}$ describes the coupling coefficient between the power flow into the cavity and the power flow into the field emission electron source which is independent of the geometry and material parameters. Here $g_{c}$ is a slowly varying function of local electric field ($E_{l}$) where extensive research and field emission electron sources have found that this value saturates during a breakdown at 10 GV/m which corresponds to a $g_{c}$=0.22. Fig~\ref{Sc} shows the magnitude of S$_c$ along a cross-section of the cavity presented in this paper

Fig.~\ref{Sc} shows that the maximum of $S_{c}$ is not located at the beam iris. A. Grudiev et al. showed in their experiments that a structure with a similar location of the peak of Sc was able to show an improvement in the maximum achievable peak surface electric fields when compared to the structures analyzed in their report, implying the location of Sc is a contributing factor to the maximum achievable fields in the structure. This means that the location of the maximum of the surface fields due to the change in the location of the modified Poynting vector could be another mechanism for dominating the BDR other than pulse heating.

In summary, we designed an optimal geometry to minimize pulse heating on the waveguide coupler and to minimize the peak surface electric field on the irises. This resulted in a maximum modified Poynting vector located further from the beam axis, which may have been favorable in reducing the breakdown rate. The single-cell cavity geometry was shown to sustain remarkably high peak fields in excess of 450 MV/m. This structure was able to perform with comparable fields that are achievable at X-band. This C-band structure has a larger beam iris aperture than its X-band counterpart and therefore is attractive for use in high-charge high-gradient accelerators for proton and electron machines. This experiment also confirmed that the structure made from a CuAg alloy can withstand significantly higher fields than its Cu counterpart for the same breakdown rates. In future works we will test two different methodologies for directly measuring the field to validate the fields calculated here. The first is to use direct probe measurements inside the cavity or measure the energy of dark current and the bremsstrahlung radiation due to dark current/field emission sources. These secondary measurements will also be able to determine the beam loading due to dark current/field emission effects to determine if this is a significant mechanism for determining the breakdown statistics.

\noindent \textbf{Acknowledgment.} LANL was supported by Los Alamos National Laboratory’s LDRD Program and Technology Evaluation and Development (TED) funds. SLAC was supported by the Department of Energy Contract No. DEAC02-76SF00515. The Authors would like to also acknowledge the excellent support and mentoring and providing excellent undergraduate and graduate students through the ASET in particular Jem Jevarjian and Sergey V. Baryshev.
\newline
\noindent \textbf{Data Availability}
The data that support the findings of this study are available from the corresponding author upon reasonable request.

\bibliography{Cband}

\begin{thebibliography}{28}%
\makeatletter
\providecommand \@ifxundefined [1]{%
 \@ifx{#1\undefined}
}%
\providecommand \@ifnum [1]{%
 \ifnum #1\expandafter \@firstoftwo
 \else \expandafter \@secondoftwo
 \fi
}%
\providecommand \@ifx [1]{%
 \ifx #1\expandafter \@firstoftwo
 \else \expandafter \@secondoftwo
 \fi
}%
\providecommand \natexlab [1]{#1}%
\providecommand \enquote  [1]{``#1''}%
\providecommand \bibnamefont  [1]{#1}%
\providecommand \bibfnamefont [1]{#1}%
\providecommand \citenamefont [1]{#1}%
\providecommand \href@noop [0]{\@secondoftwo}%
\providecommand \href [0]{\begingroup \@sanitize@url \@href}%
\providecommand \@href[1]{\@@startlink{#1}\@@href}%
\providecommand \@@href[1]{\endgroup#1\@@endlink}%
\providecommand \@sanitize@url [0]{\catcode `\\12\catcode `\$12\catcode
  `\&12\catcode `\#12\catcode `\^12\catcode `\_12\catcode `\%12\relax}%
\providecommand \@@startlink[1]{}%
\providecommand \@@endlink[0]{}%
\providecommand \url  [0]{\begingroup\@sanitize@url \@url }%
\providecommand \@url [1]{\endgroup\@href {#1}{\urlprefix }}%
\providecommand \urlprefix  [0]{URL }%
\providecommand \Eprint [0]{\href }%
\providecommand \doibase [0]{http://dx.doi.org/}%
\providecommand \selectlanguage [0]{\@gobble}%
\providecommand \bibinfo  [0]{\@secondoftwo}%
\providecommand \bibfield  [0]{\@secondoftwo}%
\providecommand \translation [1]{[#1]}%
\providecommand \BibitemOpen [0]{}%
\providecommand \bibitemStop [0]{}%
\providecommand \bibitemNoStop [0]{.\EOS\space}%
\providecommand \EOS [0]{\spacefactor3000\relax}%
\providecommand \BibitemShut  [1]{\csname bibitem#1\endcsname}%
\let\auto@bib@innerbib\@empty
\bibitem [{\citenamefont {Emma}\ \emph {et~al.}(2010)\citenamefont {Emma},
  \citenamefont {Akre}, \citenamefont {Arthur}, \citenamefont {Bionta},
  \citenamefont {Bostedt}, \citenamefont {Bozek}, \citenamefont {Brachmann},
  \citenamefont {Bucksbaum}, \citenamefont {Coffee},\ and\ \citenamefont
  {Decker}}]{FEL}%
  \BibitemOpen
  \bibfield  {author} {\bibinfo {author} {\bibfnamefont {P.}~\bibnamefont
  {Emma}}, \bibinfo {author} {\bibfnamefont {R.}~\bibnamefont {Akre}}, \bibinfo
  {author} {\bibfnamefont {J.}~\bibnamefont {Arthur}}, \bibinfo {author}
  {\bibfnamefont {R.}~\bibnamefont {Bionta}}, \bibinfo {author} {\bibfnamefont
  {C.}~\bibnamefont {Bostedt}}, \bibinfo {author} {\bibfnamefont
  {J.}~\bibnamefont {Bozek}}, \bibinfo {author} {\bibfnamefont
  {A.}~\bibnamefont {Brachmann}}, \bibinfo {author} {\bibfnamefont
  {P.}~\bibnamefont {Bucksbaum}}, \bibinfo {author} {\bibfnamefont
  {R.}~\bibnamefont {Coffee}}, \ and\ \bibinfo {author} {\bibfnamefont
  {Y.}~\bibnamefont {Decker}, \bibfnamefont {F.J.and~Ding}},\ }\href {\doibase
  10.1038/nphoton.2010.176} {\bibfield  {journal} {\bibinfo  {journal} {Nat.
  Photonics}\ }\textbf {\bibinfo {volume} {4}},\ \bibinfo {pages} {641}
  (\bibinfo {year} {2010})}\BibitemShut {NoStop}%
\bibitem [{\citenamefont {Posen}\ \emph {et~al.}(2021)\citenamefont {Posen},
  \citenamefont {Cravatta}, \citenamefont {Checchin}, \citenamefont {Aderhold},
  \citenamefont {Adolphsen}, \citenamefont {Arkan}, \citenamefont {Bafia},
  \citenamefont {Benwell}, \citenamefont {Bice}, \citenamefont {Chase},\ and\
  \citenamefont {Contreras-Martinez}}]{lclsHE}%
  \BibitemOpen
  \bibfield  {author} {\bibinfo {author} {\bibfnamefont {S.}~\bibnamefont
  {Posen}}, \bibinfo {author} {\bibfnamefont {A.}~\bibnamefont {Cravatta}},
  \bibinfo {author} {\bibfnamefont {M.}~\bibnamefont {Checchin}}, \bibinfo
  {author} {\bibfnamefont {S.}~\bibnamefont {Aderhold}}, \bibinfo {author}
  {\bibfnamefont {C.}~\bibnamefont {Adolphsen}}, \bibinfo {author}
  {\bibfnamefont {T.}~\bibnamefont {Arkan}}, \bibinfo {author} {\bibfnamefont
  {D.}~\bibnamefont {Bafia}}, \bibinfo {author} {\bibfnamefont
  {A.}~\bibnamefont {Benwell}}, \bibinfo {author} {\bibfnamefont
  {D.}~\bibnamefont {Bice}}, \bibinfo {author} {\bibfnamefont {B.}~\bibnamefont
  {Chase}}, \ and\ \bibinfo {author} {\bibfnamefont {C.}~\bibnamefont
  {Contreras-Martinez}},\ }\href {\doibase arXiv:2110.14580} {\bibfield
  {journal} {\bibinfo  {journal} {arXiv preprint}\ } (\bibinfo {year} {2021}),\
  arXiv:2110.14580}\BibitemShut {NoStop}%
\bibitem [{\citenamefont {Shiltsev}\ and\ \citenamefont
  {Zimmermann}(2021)}]{Nextgenlinac}%
  \BibitemOpen
  \bibfield  {author} {\bibinfo {author} {\bibfnamefont {V.}~\bibnamefont
  {Shiltsev}}\ and\ \bibinfo {author} {\bibfnamefont {F.}~\bibnamefont
  {Zimmermann}},\ }\href {\doibase 10.1103/RevModPhys.93.015006} {\bibfield
  {journal} {\bibinfo  {journal} {Reviews of Modern Physics}\ }\textbf
  {\bibinfo {volume} {93}},\ \bibinfo {pages} {015006} (\bibinfo {year}
  {2021})}\BibitemShut {NoStop}%
\bibitem [{\citenamefont {Nanni}\ \emph {et~al.}(2022)\citenamefont {Nanni},
  \citenamefont {Breidenbach}, \citenamefont {Vernieri}, \citenamefont
  {Belomestnykh}, \citenamefont {Bhat},\ and\ \citenamefont {et~al.}}]{C3}%
  \BibitemOpen
  \bibfield  {author} {\bibinfo {author} {\bibfnamefont {E.~A.}\ \bibnamefont
  {Nanni}}, \bibinfo {author} {\bibfnamefont {M.}~\bibnamefont {Breidenbach}},
  \bibinfo {author} {\bibfnamefont {C.}~\bibnamefont {Vernieri}}, \bibinfo
  {author} {\bibfnamefont {S.}~\bibnamefont {Belomestnykh}}, \bibinfo {author}
  {\bibfnamefont {P.}~\bibnamefont {Bhat}}, \ and\ \bibinfo {author}
  {\bibfnamefont {N.}~\bibnamefont {et~al.}},\ }\href {\doibase
  arXiv:2203.09076} {\bibfield  {journal} {\bibinfo  {journal} {arXiv
  preprint}\ } (\bibinfo {year} {2022}),\ arXiv:2203.09076}\BibitemShut
  {NoStop}%
\bibitem [{\citenamefont {Bai}\ \emph {et~al.}(2021)\citenamefont {Bai},
  \citenamefont {Barklow}, \citenamefont {Bartoldus}, \citenamefont
  {Breidenbach}, \citenamefont {Grenier}, \citenamefont {Huang}, \citenamefont
  {Kagan}, \citenamefont {Lewellen}, \citenamefont {Li}, \citenamefont
  {Markiewicz} \emph {et~al.}}]{bai2021c}%
  \BibitemOpen
  \bibfield  {author} {\bibinfo {author} {\bibfnamefont {M.}~\bibnamefont
  {Bai}}, \bibinfo {author} {\bibfnamefont {T.}~\bibnamefont {Barklow}},
  \bibinfo {author} {\bibfnamefont {R.}~\bibnamefont {Bartoldus}}, \bibinfo
  {author} {\bibfnamefont {M.}~\bibnamefont {Breidenbach}}, \bibinfo {author}
  {\bibfnamefont {P.}~\bibnamefont {Grenier}}, \bibinfo {author} {\bibfnamefont
  {Z.}~\bibnamefont {Huang}}, \bibinfo {author} {\bibfnamefont
  {M.}~\bibnamefont {Kagan}}, \bibinfo {author} {\bibfnamefont
  {J.}~\bibnamefont {Lewellen}}, \bibinfo {author} {\bibfnamefont
  {Z.}~\bibnamefont {Li}}, \bibinfo {author} {\bibfnamefont {T.~W.}\
  \bibnamefont {Markiewicz}},  \emph {et~al.},\ }\href@noop {} {\bibfield
  {journal} {\bibinfo  {journal} {arXiv preprint arXiv:2110.15800}\ } (\bibinfo
  {year} {2021})}\BibitemShut {NoStop}%
\bibitem [{\citenamefont {Lu}\ \emph {et~al.}(2021)\citenamefont {Lu},
  \citenamefont {Li}, \citenamefont {Dolgashev}, \citenamefont {Bowden},
  \citenamefont {Sy}, \citenamefont {Tantawi},\ and\ \citenamefont
  {Nanni}}]{SbandXlu}%
  \BibitemOpen
  \bibfield  {author} {\bibinfo {author} {\bibfnamefont {X.}~\bibnamefont
  {Lu}}, \bibinfo {author} {\bibfnamefont {Z.}~\bibnamefont {Li}}, \bibinfo
  {author} {\bibfnamefont {V.}~\bibnamefont {Dolgashev}}, \bibinfo {author}
  {\bibfnamefont {G.}~\bibnamefont {Bowden}}, \bibinfo {author} {\bibfnamefont
  {A.}~\bibnamefont {Sy}}, \bibinfo {author} {\bibfnamefont {S.}~\bibnamefont
  {Tantawi}}, \ and\ \bibinfo {author} {\bibfnamefont {E.~A.}\ \bibnamefont
  {Nanni}},\ }\href {\doibase 10.1063/5.0035331} {\bibfield  {journal}
  {\bibinfo  {journal} {Review of Scientific Instruments}\ }\textbf {\bibinfo
  {volume} {92}},\ \bibinfo {pages} {024705} (\bibinfo {year}
  {2021})}\BibitemShut {NoStop}%
\bibitem [{\citenamefont {V.}(2014)}]{Medlinac}%
  \BibitemOpen
  \bibfield  {author} {\bibinfo {author} {\bibfnamefont {M.}~\bibnamefont
  {V.}},\ }\href {\doibase 10.1038/508133a} {\bibfield  {journal} {\bibinfo
  {journal} {Nature}\ }\textbf {\bibinfo {volume} {508}},\ \bibinfo {pages}
  {133} (\bibinfo {year} {2014})}\BibitemShut {NoStop}%
\bibitem [{\citenamefont {Kutsaev}\ \emph {et~al.}(2015)\citenamefont
  {Kutsaev}, \citenamefont {Agustsson}, \citenamefont {Arodzero}, \citenamefont
  {Faillace}, \citenamefont {J.},\ and\ \citenamefont {V.}}]{Industry}%
  \BibitemOpen
  \bibfield  {author} {\bibinfo {author} {\bibfnamefont {S.~V.}\ \bibnamefont
  {Kutsaev}}, \bibinfo {author} {\bibfnamefont {R.}~\bibnamefont {Agustsson}},
  \bibinfo {author} {\bibfnamefont {A.}~\bibnamefont {Arodzero}}, \bibinfo
  {author} {\bibfnamefont {L.}~\bibnamefont {Faillace}}, \bibinfo {author}
  {\bibfnamefont {H.}~\bibnamefont {J.}}, \ and\ \bibinfo {author}
  {\bibfnamefont {Z.}~\bibnamefont {V.}},\ }\href@noop {} {\bibfield  {journal}
  {\bibinfo  {journal} {Proc.\ IEEE Nuclear Science Symposium and Medical
  Imaging Conference}\ } (\bibinfo {year} {2015})}\BibitemShut {NoStop}%
\bibitem [{\citenamefont {Simakov}\ \emph {et~al.}(2018)\citenamefont
  {Simakov}, \citenamefont {Dolgashev},\ and\ \citenamefont
  {Tantawi}}]{HGjenya}%
  \BibitemOpen
  \bibfield  {author} {\bibinfo {author} {\bibfnamefont {E.~I.}\ \bibnamefont
  {Simakov}}, \bibinfo {author} {\bibfnamefont {V.~A.}\ \bibnamefont
  {Dolgashev}}, \ and\ \bibinfo {author} {\bibfnamefont {S.~G.}\ \bibnamefont
  {Tantawi}},\ }\href {\doibase 10.1016/j.nima.2018.02.085} {\bibfield
  {journal} {\bibinfo  {journal} {Nuclear Instruments and Methods in Physics
  Research Section A.}\ }\textbf {\bibinfo {volume} {907}},\ \bibinfo {pages}
  {221} (\bibinfo {year} {2018})}\BibitemShut {NoStop}%
\bibitem [{\citenamefont {Tantawi}\ \emph {et~al.}(2020)\citenamefont
  {Tantawi}, \citenamefont {Nasr}, \citenamefont {Li}, \citenamefont
  {Limborg},\ and\ \citenamefont {Borchard}}]{SamiDC}%
  \BibitemOpen
  \bibfield  {author} {\bibinfo {author} {\bibfnamefont {S.}~\bibnamefont
  {Tantawi}}, \bibinfo {author} {\bibfnamefont {M.}~\bibnamefont {Nasr}},
  \bibinfo {author} {\bibfnamefont {Z.}~\bibnamefont {Li}}, \bibinfo {author}
  {\bibfnamefont {C.}~\bibnamefont {Limborg}}, \ and\ \bibinfo {author}
  {\bibfnamefont {P.}~\bibnamefont {Borchard}},\ }\href {\doibase
  10.1103/PhysRevAccelBeams.23.092001} {\bibfield  {journal} {\bibinfo
  {journal} {PRAB.}\ }\textbf {\bibinfo {volume} {23}},\ \bibinfo {pages}
  {092001} (\bibinfo {year} {2020})}\BibitemShut {NoStop}%
\bibitem [{\citenamefont {Nasr}\ \emph {et~al.}(2019)\citenamefont {Nasr},
  \citenamefont {Welander}, \citenamefont {Li},\ and\ \citenamefont
  {Tantawi}}]{SRFPF}%
  \BibitemOpen
  \bibfield  {author} {\bibinfo {author} {\bibfnamefont {M.~H.}\ \bibnamefont
  {Nasr}}, \bibinfo {author} {\bibfnamefont {P.~B.}\ \bibnamefont {Welander}},
  \bibinfo {author} {\bibfnamefont {Z.}~\bibnamefont {Li}}, \ and\ \bibinfo
  {author} {\bibfnamefont {S.}~\bibnamefont {Tantawi}},\ }\href {\doibase
  10.18429/JACoW-IPAC2019-WEPRB090} {\bibfield  {journal} {\bibinfo  {journal}
  {Proc.\ 10th Int. Particle Accelerator Conf.}\ } (\bibinfo {year} {2019}),\
  10.18429/JACoW-IPAC2019-WEPRB090}\BibitemShut {NoStop}%
\bibitem [{\citenamefont {Dolgashev}\ \emph {et~al.}(2010)\citenamefont
  {Dolgashev}, \citenamefont {Tantawi}, \citenamefont {Higashi},\ and\
  \citenamefont {Spataro}}]{GeoValery}%
  \BibitemOpen
  \bibfield  {author} {\bibinfo {author} {\bibfnamefont {V.}~\bibnamefont
  {Dolgashev}}, \bibinfo {author} {\bibfnamefont {S.}~\bibnamefont {Tantawi}},
  \bibinfo {author} {\bibfnamefont {Y.}~\bibnamefont {Higashi}}, \ and\
  \bibinfo {author} {\bibfnamefont {B.}~\bibnamefont {Spataro}},\ }\href
  {\doibase 10.1063/1.3505339} {\bibfield  {journal} {\bibinfo  {journal}
  {Applied Physics Letters}\ }\textbf {\bibinfo {volume} {97}},\ \bibinfo
  {pages} {171501} (\bibinfo {year} {2010})}\BibitemShut {NoStop}%
\bibitem [{\citenamefont {Laurent}\ \emph {et~al.}(2011)\citenamefont
  {Laurent}, \citenamefont {Tantawi}, \citenamefont {Dolgashev}, \citenamefont
  {Nantista}, \citenamefont {Higashi}, \citenamefont {Aicheler}, \citenamefont
  {Heikkinen},\ and\ \citenamefont {Wuensch}}]{PHValery}%
  \BibitemOpen
  \bibfield  {author} {\bibinfo {author} {\bibfnamefont {L.}~\bibnamefont
  {Laurent}}, \bibinfo {author} {\bibfnamefont {S.}~\bibnamefont {Tantawi}},
  \bibinfo {author} {\bibfnamefont {V.}~\bibnamefont {Dolgashev}}, \bibinfo
  {author} {\bibfnamefont {C.}~\bibnamefont {Nantista}}, \bibinfo {author}
  {\bibfnamefont {Y.}~\bibnamefont {Higashi}}, \bibinfo {author} {\bibfnamefont
  {M.}~\bibnamefont {Aicheler}}, \bibinfo {author} {\bibfnamefont
  {S.}~\bibnamefont {Heikkinen}}, \ and\ \bibinfo {author} {\bibfnamefont
  {W.}~\bibnamefont {Wuensch}},\ }\href {\doibase
  10.1103/PhysRevSTAB.14.041001} {\bibfield  {journal} {\bibinfo  {journal}
  {PRAB.}\ }\textbf {\bibinfo {volume} {14}},\ \bibinfo {pages} {041001}
  (\bibinfo {year} {2011})}\BibitemShut {NoStop}%
\bibitem [{\citenamefont {Cahill}\ \emph {et~al.}(2018)\citenamefont {Cahill},
  \citenamefont {Rosenzweig}, \citenamefont {Dolgashev}, \citenamefont
  {Tantawi},\ and\ \citenamefont {Weathersby}}]{SRFValery}%
  \BibitemOpen
  \bibfield  {author} {\bibinfo {author} {\bibfnamefont {A.~D.}\ \bibnamefont
  {Cahill}}, \bibinfo {author} {\bibfnamefont {J.~B.}\ \bibnamefont
  {Rosenzweig}}, \bibinfo {author} {\bibfnamefont {V.~A.}\ \bibnamefont
  {Dolgashev}}, \bibinfo {author} {\bibfnamefont {S.~G.}\ \bibnamefont
  {Tantawi}}, \ and\ \bibinfo {author} {\bibfnamefont {S.}~\bibnamefont
  {Weathersby}},\ }\href {\doibase 10.1103/PhysRevAccelBeams.21.102002}
  {\bibfield  {journal} {\bibinfo  {journal} {PRAB.}\ }\textbf {\bibinfo
  {volume} {21}},\ \bibinfo {pages} {102002} (\bibinfo {year}
  {2018})}\BibitemShut {NoStop}%
\bibitem [{\citenamefont {Othman}\ \emph {et~al.}(2020)\citenamefont {Othman},
  \citenamefont {Picard}, \citenamefont {Schaub}, \citenamefont {Dolgashev},
  \citenamefont {Lewis}, \citenamefont {Neilson}, \citenamefont {Haase},
  \citenamefont {Jawla}, \citenamefont {Spataro}, \citenamefont {Temkin} \emph
  {et~al.}}]{othman2020experimental}%
  \BibitemOpen
  \bibfield  {author} {\bibinfo {author} {\bibfnamefont {M.~A.}\ \bibnamefont
  {Othman}}, \bibinfo {author} {\bibfnamefont {J.}~\bibnamefont {Picard}},
  \bibinfo {author} {\bibfnamefont {S.}~\bibnamefont {Schaub}}, \bibinfo
  {author} {\bibfnamefont {V.~A.}\ \bibnamefont {Dolgashev}}, \bibinfo {author}
  {\bibfnamefont {S.~M.}\ \bibnamefont {Lewis}}, \bibinfo {author}
  {\bibfnamefont {J.}~\bibnamefont {Neilson}}, \bibinfo {author} {\bibfnamefont
  {A.}~\bibnamefont {Haase}}, \bibinfo {author} {\bibfnamefont
  {S.}~\bibnamefont {Jawla}}, \bibinfo {author} {\bibfnamefont
  {B.}~\bibnamefont {Spataro}}, \bibinfo {author} {\bibfnamefont {R.~J.}\
  \bibnamefont {Temkin}},  \emph {et~al.},\ }\href@noop {} {\bibfield
  {journal} {\bibinfo  {journal} {Applied Physics Letters}\ }\textbf {\bibinfo
  {volume} {117}},\ \bibinfo {pages} {073502} (\bibinfo {year}
  {2020})}\BibitemShut {NoStop}%
\bibitem [{\citenamefont {Shao}(2018)}]{Jiahang}%
  \BibitemOpen
  \bibfield  {author} {\bibinfo {author} {\bibfnamefont {J.}~\bibnamefont
  {Shao}},\ }\href {\doibase 10.1007/978-981-10-7926-9} {\bibfield  {journal}
  {\bibinfo  {journal} {Springer}\ } (\bibinfo {year} {2018}),\
  10.1007/978-981-10-7926-9}\BibitemShut {NoStop}%
\bibitem [{\citenamefont {Loehl}\ \emph {et~al.}(2013)\citenamefont {Loehl},
  \citenamefont {Alex}, \citenamefont {Blumer}, \citenamefont {Bopp},
  \citenamefont {Braun}, \citenamefont {Citterio}, \citenamefont {Ellenberger},
  \citenamefont {Fitze}, \citenamefont {Joehri}, \citenamefont {Kleeb},
  \citenamefont {Paly}, \citenamefont {Raguin}, \citenamefont {Schulz},\ and\
  \citenamefont {Zennaro}}]{SwissFEL}%
  \BibitemOpen
  \bibfield  {author} {\bibinfo {author} {\bibfnamefont {F.}~\bibnamefont
  {Loehl}}, \bibinfo {author} {\bibfnamefont {J.}~\bibnamefont {Alex}},
  \bibinfo {author} {\bibfnamefont {H.}~\bibnamefont {Blumer}}, \bibinfo
  {author} {\bibfnamefont {M.}~\bibnamefont {Bopp}}, \bibinfo {author}
  {\bibfnamefont {H.}~\bibnamefont {Braun}}, \bibinfo {author} {\bibfnamefont
  {A.}~\bibnamefont {Citterio}}, \bibinfo {author} {\bibfnamefont
  {U.}~\bibnamefont {Ellenberger}}, \bibinfo {author} {\bibfnamefont
  {H.}~\bibnamefont {Fitze}}, \bibinfo {author} {\bibfnamefont
  {H.}~\bibnamefont {Joehri}}, \bibinfo {author} {\bibfnamefont
  {T.}~\bibnamefont {Kleeb}}, \bibinfo {author} {\bibfnamefont
  {L.}~\bibnamefont {Paly}}, \bibinfo {author} {\bibfnamefont {J.}~\bibnamefont
  {Raguin}}, \bibinfo {author} {\bibfnamefont {L.}~\bibnamefont {Schulz}}, \
  and\ \bibinfo {author} {\bibfnamefont {R.}~\bibnamefont {Zennaro}},\
  }\href@noop {} {\bibfield  {journal} {\bibinfo  {journal} {Proc.\ FEL2013}\ }
  (\bibinfo {year} {2013})}\BibitemShut {NoStop}%
\bibitem [{\citenamefont {Fang}\ \emph {et~al.}(2016)\citenamefont {Fang},
  \citenamefont {Gu}, \citenamefont {Sheng}, \citenamefont {Wang},
  \citenamefont {Tong}, \citenamefont {Chen}, \citenamefont {Zhong},
  \citenamefont {Tan}, \citenamefont {Lin}, \citenamefont {Chen},\ and\
  \citenamefont {Zhao}}]{SINAP}%
  \BibitemOpen
  \bibfield  {author} {\bibinfo {author} {\bibfnamefont {W.}~\bibnamefont
  {Fang}}, \bibinfo {author} {\bibfnamefont {Q.}~\bibnamefont {Gu}}, \bibinfo
  {author} {\bibfnamefont {X.}~\bibnamefont {Sheng}}, \bibinfo {author}
  {\bibfnamefont {C.}~\bibnamefont {Wang}}, \bibinfo {author} {\bibfnamefont
  {D.}~\bibnamefont {Tong}}, \bibinfo {author} {\bibfnamefont {L.}~\bibnamefont
  {Chen}}, \bibinfo {author} {\bibfnamefont {S.}~\bibnamefont {Zhong}},
  \bibinfo {author} {\bibfnamefont {J.}~\bibnamefont {Tan}}, \bibinfo {author}
  {\bibfnamefont {G.}~\bibnamefont {Lin}}, \bibinfo {author} {\bibfnamefont
  {Z.}~\bibnamefont {Chen}}, \ and\ \bibinfo {author} {\bibfnamefont
  {Z.}~\bibnamefont {Zhao}},\ }\href {\doibase 10.1016/j.nima.2016.03.101}
  {\bibfield  {journal} {\bibinfo  {journal} {Nuclear Instruments and Methods
  in Physics Research Section A.}\ }\textbf {\bibinfo {volume} {823}},\
  \bibinfo {pages} {91} (\bibinfo {year} {2016})}\BibitemShut {NoStop}%
\bibitem [{\citenamefont {Alesini}\ \emph {et~al.}(2020)\citenamefont
  {Alesini}, \citenamefont {Bellaveglia}, \citenamefont {Cardelli},
  \citenamefont {Di~Raddo}, \citenamefont {Gallo}, \citenamefont {Lollo},
  \citenamefont {Piersanti}, \citenamefont {Variola}, \citenamefont {Palumbo},
  \citenamefont {Poletto},\ and\ \citenamefont {Favaron}}]{SPARC}%
  \BibitemOpen
  \bibfield  {author} {\bibinfo {author} {\bibfnamefont {D.}~\bibnamefont
  {Alesini}}, \bibinfo {author} {\bibfnamefont {M.}~\bibnamefont
  {Bellaveglia}}, \bibinfo {author} {\bibfnamefont {F.}~\bibnamefont
  {Cardelli}}, \bibinfo {author} {\bibfnamefont {R.}~\bibnamefont {Di~Raddo}},
  \bibinfo {author} {\bibfnamefont {A.}~\bibnamefont {Gallo}}, \bibinfo
  {author} {\bibfnamefont {V.}~\bibnamefont {Lollo}}, \bibinfo {author}
  {\bibfnamefont {L.}~\bibnamefont {Piersanti}}, \bibinfo {author}
  {\bibfnamefont {A.}~\bibnamefont {Variola}}, \bibinfo {author} {\bibfnamefont
  {L.}~\bibnamefont {Palumbo}}, \bibinfo {author} {\bibfnamefont
  {F.}~\bibnamefont {Poletto}}, \ and\ \bibinfo {author} {\bibfnamefont
  {P.}~\bibnamefont {Favaron}},\ }\href {\doibase
  10.1103/PhysRevAccelBeams.23.042001} {\bibfield  {journal} {\bibinfo
  {journal} {PRAB.}\ }\textbf {\bibinfo {volume} {23}},\ \bibinfo {pages}
  {042001} (\bibinfo {year} {2020})}\BibitemShut {NoStop}%
\bibitem [{\citenamefont {Shintake}(2019)}]{Spring8}%
  \BibitemOpen
  \bibfield  {author} {\bibinfo {author} {\bibfnamefont {T.}~\bibnamefont
  {Shintake}},\ }\href {\doibase 10.18429/JACoW-IPAC2019-WEPRB090} {\bibfield
  {journal} {\bibinfo  {journal} {Proc.\ 10th Int. Particle Accelerator Conf.}\
  } (\bibinfo {year} {2019}),\ 10.18429/JACoW-IPAC2019-WEPRB090}\BibitemShut
  {NoStop}%
\bibitem [{\citenamefont {Yang}\ \emph {et~al.}(2010)\citenamefont {Yang},
  \citenamefont {Kim}, \citenamefont {Park}, \citenamefont {Cho}, \citenamefont
  {Namkung}, \citenamefont {Oh}, \citenamefont {Lee},\ and\ \citenamefont
  {Chung}}]{KNFRI}%
  \BibitemOpen
  \bibfield  {author} {\bibinfo {author} {\bibfnamefont {H.}~\bibnamefont
  {Yang}}, \bibinfo {author} {\bibfnamefont {S.}~\bibnamefont {Kim}}, \bibinfo
  {author} {\bibfnamefont {S.}~\bibnamefont {Park}}, \bibinfo {author}
  {\bibfnamefont {M.}~\bibnamefont {Cho}}, \bibinfo {author} {\bibfnamefont
  {W.}~\bibnamefont {Namkung}}, \bibinfo {author} {\bibfnamefont
  {J.}~\bibnamefont {Oh}}, \bibinfo {author} {\bibfnamefont {K.}~\bibnamefont
  {Lee}}, \ and\ \bibinfo {author} {\bibfnamefont {K.}~\bibnamefont {Chung}},\
  }\href@noop {} {\bibfield  {journal} {\bibinfo  {journal} {Proc.\ Linear
  Accelerator Conf. LINAC2010}\ } (\bibinfo {year} {2010})}\BibitemShut
  {NoStop}%
\bibitem [{\citenamefont {Wu}\ \emph {et~al.}(2017)\citenamefont {Wu},
  \citenamefont {Shi}, \citenamefont {Chen}, \citenamefont {Shao},
  \citenamefont {Abe}, \citenamefont {Higo}, \citenamefont {Matsumoto},\ and\
  \citenamefont {Wuensch}}]{CERNKEK}%
  \BibitemOpen
  \bibfield  {author} {\bibinfo {author} {\bibfnamefont {X.}~\bibnamefont
  {Wu}}, \bibinfo {author} {\bibfnamefont {J.}~\bibnamefont {Shi}}, \bibinfo
  {author} {\bibfnamefont {H.}~\bibnamefont {Chen}}, \bibinfo {author}
  {\bibfnamefont {J.}~\bibnamefont {Shao}}, \bibinfo {author} {\bibfnamefont
  {T.}~\bibnamefont {Abe}}, \bibinfo {author} {\bibfnamefont {T.}~\bibnamefont
  {Higo}}, \bibinfo {author} {\bibfnamefont {S.}~\bibnamefont {Matsumoto}}, \
  and\ \bibinfo {author} {\bibfnamefont {W.}~\bibnamefont {Wuensch}},\ }\href
  {\doibase 10.1103/PhysRevAccelBeams.20.052001} {\bibfield  {journal}
  {\bibinfo  {journal} {PRAB.}\ }\textbf {\bibinfo {volume} {20}},\ \bibinfo
  {pages} {052001} (\bibinfo {year} {2017})}\BibitemShut {NoStop}%
\bibitem [{\citenamefont {Vnuchenko}\ \emph {et~al.}(2020)\citenamefont
  {Vnuchenko}, \citenamefont {Pereira}, \citenamefont {Martinez}, \citenamefont
  {Benedetti}, \citenamefont {Lasheras}, \citenamefont {Garlasch},
  \citenamefont {Grudiev}, \citenamefont {McMonagle}, \citenamefont {Pitman},
  \citenamefont {Syratchev}, \citenamefont {Timmins},\ and\ \citenamefont
  {et~al.}}]{CERN}%
  \BibitemOpen
  \bibfield  {author} {\bibinfo {author} {\bibfnamefont {A.}~\bibnamefont
  {Vnuchenko}}, \bibinfo {author} {\bibfnamefont {D.}~\bibnamefont {Pereira}},
  \bibinfo {author} {\bibfnamefont {B.}~\bibnamefont {Martinez}}, \bibinfo
  {author} {\bibfnamefont {S.}~\bibnamefont {Benedetti}}, \bibinfo {author}
  {\bibfnamefont {N.}~\bibnamefont {Lasheras}}, \bibinfo {author}
  {\bibfnamefont {M.}~\bibnamefont {Garlasch}}, \bibinfo {author}
  {\bibfnamefont {A.}~\bibnamefont {Grudiev}}, \bibinfo {author} {\bibfnamefont
  {G.}~\bibnamefont {McMonagle}}, \bibinfo {author} {\bibfnamefont
  {S.}~\bibnamefont {Pitman}}, \bibinfo {author} {\bibfnamefont
  {I.}~\bibnamefont {Syratchev}}, \bibinfo {author} {\bibfnamefont
  {M.}~\bibnamefont {Timmins}}, \ and\ \bibinfo {author} {\bibnamefont
  {et~al.}},\ }\href {\doibase 10.1103/PhysRevAccelBeams.23.084801} {\bibfield
  {journal} {\bibinfo  {journal} {PRAB.}\ }\textbf {\bibinfo {volume} {23}},\
  \bibinfo {pages} {084801} (\bibinfo {year} {2020})}\BibitemShut {NoStop}%
\bibitem [{\citenamefont {Wang}\ \emph {et~al.}(2022)\citenamefont {Wang},
  \citenamefont {Simakov},\ and\ \citenamefont {Perez}}]{LANLThm}%
  \BibitemOpen
  \bibfield  {author} {\bibinfo {author} {\bibfnamefont {G.}~\bibnamefont
  {Wang}}, \bibinfo {author} {\bibfnamefont {E.~I.}\ \bibnamefont {Simakov}}, \
  and\ \bibinfo {author} {\bibfnamefont {D.}~\bibnamefont {Perez}},\ }\href
  {\doibase 10.1063/5.0084266} {\bibfield  {journal} {\bibinfo  {journal} {App.
  Phys. Lett.}\ }\textbf {\bibinfo {volume} {120}},\ \bibinfo {pages} {134101}
  (\bibinfo {year} {2022})}\BibitemShut {NoStop}%
\bibitem [{\citenamefont {Gorelov}\ \emph {et~al.}(2021)\citenamefont
  {Gorelov}, \citenamefont {Fleming}, \citenamefont {Lawrence}, \citenamefont
  {Lewellen}, \citenamefont {Middendorf}, \citenamefont {Perez}, \citenamefont
  {Simakov},\ and\ \citenamefont {Tajiman}}]{CERF}%
  \BibitemOpen
  \bibfield  {author} {\bibinfo {author} {\bibfnamefont {D.}~\bibnamefont
  {Gorelov}}, \bibinfo {author} {\bibfnamefont {R.~L.}\ \bibnamefont
  {Fleming}}, \bibinfo {author} {\bibfnamefont {S.~K.}\ \bibnamefont
  {Lawrence}}, \bibinfo {author} {\bibfnamefont {J.~W.}\ \bibnamefont
  {Lewellen}}, \bibinfo {author} {\bibfnamefont {M.~E.}\ \bibnamefont
  {Middendorf}}, \bibinfo {author} {\bibfnamefont {M.~E.}\ \bibnamefont
  {Perez}, \bibfnamefont {D.~Schneider}}, \bibinfo {author} {\bibfnamefont
  {E.~I.}\ \bibnamefont {Simakov}}, \ and\ \bibinfo {author} {\bibfnamefont
  {T.}~\bibnamefont {Tajiman}},\ }\href@noop {} {\bibfield  {journal} {\bibinfo
   {journal} {Proc.\ 12th Int. Particle Acc. Conf.}\ } (\bibinfo {year}
  {2021})}\BibitemShut {NoStop}%
\bibitem [{\citenamefont {Schneider}\ \emph {et~al.}(2021)\citenamefont
  {Schneider}, \citenamefont {Fleming}, \citenamefont {Simakov}, \citenamefont
  {Gorelov}, \citenamefont {Lewellen}, \citenamefont {Jevarjian},\ and\
  \citenamefont {Baryshev}}]{FEbreak}%
  \BibitemOpen
  \bibfield  {author} {\bibinfo {author} {\bibfnamefont {M.}~\bibnamefont
  {Schneider}}, \bibinfo {author} {\bibfnamefont {R.}~\bibnamefont {Fleming}},
  \bibinfo {author} {\bibfnamefont {E.}~\bibnamefont {Simakov}}, \bibinfo
  {author} {\bibfnamefont {D.}~\bibnamefont {Gorelov}}, \bibinfo {author}
  {\bibfnamefont {J.}~\bibnamefont {Lewellen}}, \bibinfo {author}
  {\bibfnamefont {E.}~\bibnamefont {Jevarjian}}, \ and\ \bibinfo {author}
  {\bibfnamefont {S.~V.}\ \bibnamefont {Baryshev}},\ }\href {\doibase
  10.18429/JACoW-IPAC2021-THPAB138} {\bibfield  {journal} {\bibinfo  {journal}
  {Proc.\ 12th Int. Particle Acc. Conf.}\ } (\bibinfo {year} {2021}),\
  10.18429/JACoW-IPAC2021-THPAB138}\BibitemShut {NoStop}%
\bibitem [{\citenamefont {Wangler}(1998)}]{CalEa}%
  \BibitemOpen
  \bibfield  {author} {\bibinfo {author} {\bibfnamefont {T.~P.}\ \bibnamefont
  {Wangler}},\ }\href@noop {} {\bibfield  {journal} {\bibinfo  {journal} {John
  Wiley and Sons}\ } (\bibinfo {year} {1998})}\BibitemShut {NoStop}%
\bibitem [{\citenamefont {Grudiev}\ \emph {et~al.}(2009)\citenamefont
  {Grudiev}, \citenamefont {Calatroni},\ and\ \citenamefont
  {Wuensch}}]{grudiev2009new}%
  \BibitemOpen
  \bibfield  {author} {\bibinfo {author} {\bibfnamefont {A.}~\bibnamefont
  {Grudiev}}, \bibinfo {author} {\bibfnamefont {S.}~\bibnamefont {Calatroni}},
  \ and\ \bibinfo {author} {\bibfnamefont {W.}~\bibnamefont {Wuensch}},\
  }\href@noop {} {\bibfield  {journal} {\bibinfo  {journal} {Physical Review
  Special Topics-Accelerators and Beams}\ }\textbf {\bibinfo {volume} {12}},\
  \bibinfo {pages} {102001} (\bibinfo {year} {2009})}\BibitemShut {NoStop}%
\end{thebibliography}%

\end{document}